# A charge-density machine-learning workflow for computing the infrared spectrum of molecules

S. Hazra ; U. Patil ; S. Sanvito

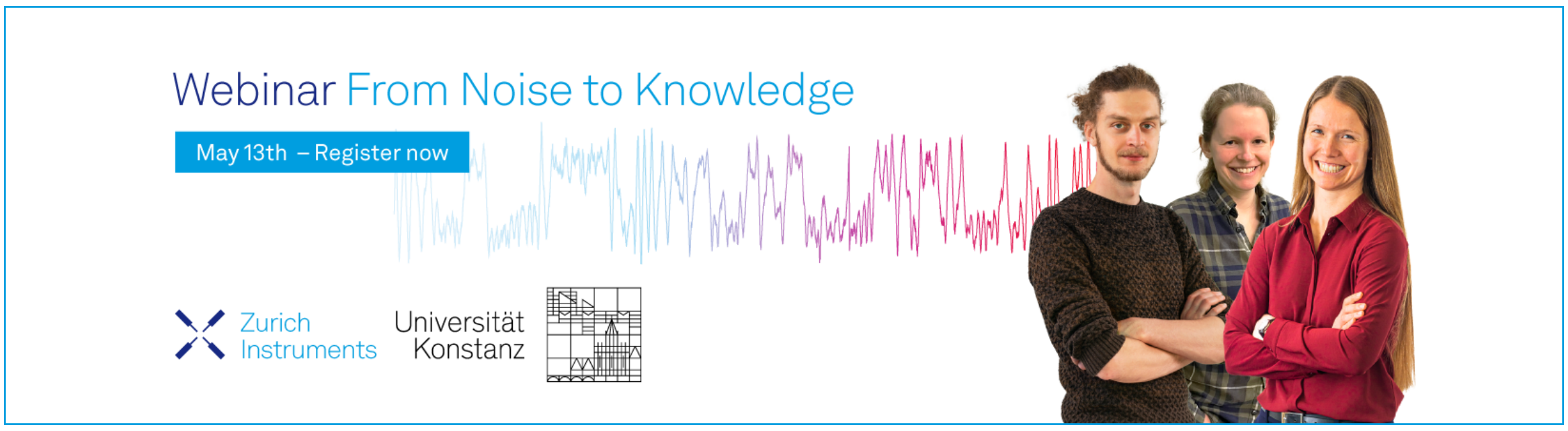

# A charge-density machine-learning workflow for computing the infrared spectrum of molecules

S. Hazra, U. Patil, and S. Sanvito[a)]

AFFILIATIONS

School of Physics and CRANN Institute, Trinity College, Dublin 2, Ireland

[a)]Author to whom correspondence should be addressed: sanvitos@tcd.ie

## ABSTRACT

We present a machine-learning workflow for the calculation of the infrared spectrum of molecules and more generally of other temperature-dependent electronic observables. The main idea is to use the Jacobi–Legendre cluster expansion to predict the real-space charge density of a converged density-functional-theory calculation. This gives us access to both energy and forces and to electronic observables such as the dipole moment or the electronic gap. Thus, the same model can simultaneously drive a molecular dynamics simulation and evaluate electronic quantities along the trajectory, namely, it has access to the same information of *ab initio* molecular dynamics. A similar approach within the framework of machine-learning force fields would require the training of multiple models, one for the molecular dynamics and others for predicting the electronic quantities. The scheme is implemented here within the numerical framework of the PySCF code and applied to the infrared spectrum of the uracil molecule in the gas phase.

## I. INTRODUCTION

Machine-learning (ML) approaches are rapidly changing computational materials science since they enable large numerical throughputs without significantly compromising on the accuracy of the *ab initio* methods they are based on.[1–4] Effectively, ML models are used as surrogates to more complex and numerically intensive schemes and they can be integrated in workflows, which ideally are less computationally demanding than those obtainable with *ab initio* methods only. Examples of ML-accelerated workflows include the prediction of crystal structures[5–7] and phase diagrams,[8–11] molecular dynamics (MD)[12,13] and the simulation of different observables either dynamical or thermodynamical.[14–17] Importantly, since the accuracy and validity of a ML model is higher close to the domain where the model has been trained on, ML accelerators are usually deployed to provide a speedup in statistical sampling, for instance, in MD or Monte Carlo schemes.

The most widely used ML algorithms in materials science are certainly ML force fields.[12,18] In most of the cases, one represents the atomic structure of molecules and solids through atomic descriptors and then constructs a numerical relation between a given structure and the total energy. When this relation is differentiable, the atomic forces follow directly from the numerical potential-energy surface, although there are also schemes to calculate directly the forces. Since one, in principle, can combine any descriptor type with any ML algorithm, there is currently a huge menu of force fields available, namely, neural networks,[19–22] kernel-based schemes,[23,24] linear regressions and cluster expansions,[25–27] and message-passing algorithms.[28,29] The choice of the specific force field for a given problem is usually determined by a large combination of materials characteristics (number of species, structural complexity, total-energy profile, etc.), conditions of use (temperature, pressure, etc.), ease of generating the training set, and numerical speed required. In any case, when a sufficient level of complexity in the representation and model is achieved, and when the training set is sufficiently wide, accuracies close to those of *ab initio* methods can be obtained. This means that a good-quality ML force field can perform MD at the same level of *ab initio* MD (AIMD). There is, however, a fundamental practical difference between MD driven by ML force fields and AIMD. In fact, with AIMD, one has direct access to all the electronic observables along the MD trajectory, information that is not available with force fields, since only energy and forces are computed. One can then remedy this drawback by constructing additional models for electronic quantities[30–33] to be run together with the force field driving the MD. This strategy, however, adds complexity to the workflow and the need to train multiple models.

An elegant and efficient solution to this problem is to use a ML model to compute the converged charge density of a density functional theory (DFT) calculation. If accurate, such charge density can then be used to evaluate ground-state observables without any self-consistent process. Since these include total energy and forces, the same model can drive MD as well, so that a single algorithm can both act as a force field and provide electronic information. In this case, together with its accuracy, the key characteristic of the model is its numerical efficiency since the inference speed of the charge density needs to be higher than that of the self-consistent cycle of an AIMD time step. This is faster than the typical convergence time of a static DFT calculation due to the fact that the initial charge density of an AIMD step is usually already close to the final one and the convergence criterion is looser. In addition for charge-density ML models, the literature is full of options differing for the density representation, the descriptors used, and the complexity of the underlining ML algorithm.[34–42]

In an attempt to construct a non-self-consistent MD scheme based on the computation of an electronic quantity, in a previous work[43] we have proposed a deep-learning model for calculating the single-particle density matrix. This, in fact, can be used to efficiently start an almost-converged DFT calculation or, for sufficiently high accuracies, to bypass completely any self-consistent step. In that instance, the single-particle density matrix was chosen over the electron density since our DFT code of choice was a local-orbital one, namely, PySCF,[44,45] which cannot be re-started directly from the charge density. The use of the density matrix, however, has several drawbacks. First, it is a covariant quantity with respect to rotations, so that any model should have covariance built in. Furthermore, the density matrix scales quadratically with the number of basis functions used to represent the operators. These two constrains make the ML model complex to construct and little transferrable.

In order to obviate these problems, we propose here an alternative workflow. We use our recently developed Jacobi–Legendre charge-density model (JLCDM)[37,39] to predict the charge density of molecules over a real space grid. Then, we use such density to perform non-self-consistent DFT calculations to extract simultaneously forces and electronic observables. The first are used to drive MD and the second to compute observables over such MD trajectories. As an example, we show how the workflow can be used to compute the infrared (IR) spectrum of molecules. Clearly, the numerical overhead of the model now scales linearly with the number of atoms since the electron density of interest is located around them. Furthermore, since the charge density at a point in space is invariant with respect to rotations, the model does not require covariance but simply invariance.

The goal of our paper is to discuss this particular workflow, highlighting both its benefits and limitations, and we have structured the discussion as follows. In Sec. II, we describe our methodology, with a special focus on how to use the real-space charge density to re-start a PySCF calculation, a feature that was not available before. In particular, we detail a newly developed algorithm to compute the matrix elements of the Hartree potential over the PySCF basis set. This is the only term in the Kohn–Sham Hamiltonian, which was not directly available in PySCF from the charge density. Next, we present a wide discussion of the main results, touching upon the accuracy of the JLCDM and of the associated energy and forces, before discussing the calculation of the IR spectrum of uracil, our molecule of choice. Finally, we conclude.

## II. THEORETICAL BACKGROUND AND METHODS

### A. Overall computational strategy

As explained in Sec. I, our aim is to demonstrate a workflow, where temperature-dependent quantities can be computed at an *ab initio* level from MD, without the need to perform self-consistent calculations for all the configurations to explore. Thus, our strategy is to construct a ML model able to predict the converged ground-state charge density, $\rho(\mathbf{r})$, from the sole knowledge of the atomic positions. The density can then be used as an input of a conventional DFT calculation, which will then return all the observables of interest. These include not only energy and forces but also electronic quantities such as the molecule dipole moment or the DFT bandgap. In fact, in contrast to ML force fields, our conceptually simple approach gives us direct access to all ground-state electronic properties without the need to construct a different model for each one of them.

The example presented here falls among this class of workflows. In fact, we will compute the infrared (IR) spectrum of molecules in the gas phase by taking the Fourier transform (FT) of the electric-dipole autocorrelation function over MD trajectories. Both the atomic forces needed for the MD and the values of the dipole moment here are computed non-self-consistently from the charge density predicted by a ML model. In particular, we use the JLCDM,[37,39] where the electron density at a generic point in space is constructed as a cluster expansion. The relevant clusters are expressed over Jacobi–Legendre polynomials and the expansion parameters are fitted through a simple linear regression.[27] The converged electron density, in principle, provides all the ground-state information of the system of interest, computed at the level of approximation offered by the chosen exchange–correlation energy. In practice, however, the way one extracts the information depends on the numerical DFT implementation used.

The most straightforward way consists in using the ML charge density as input for a single digonalization step and then computing the ground-state observables with the tools provided by the DFT code of choice. In general, this consists in solving a Roothaan–Hall-type equation[46,47] of the form

$$FC = SCE, \tag{1}$$

where $C$ is the matrix containing the coefficients of expansion of the Kohn–Sham orbitals, $E$ is the diagonal matrix of the Kohn–Sham energies, $S$ is the overlap matrix (the identity for orthonormal basis sets), and finally, $F$ is the Fock matrix. Our strategy then consists in constructing the Fock matrix from the ML charge density. In the case of a plane-wave basis set, this is quite straightforward, since one simply needs to Fourier transform the real-space density. Since this is done with a high-frequency cutoff, typically high-frequency noise is eliminated.[37] Such procedure is essentially what is done when one restarts a DFT calculation from the real-space charge density. In contrast, when one uses DFT implementations based on a local-orbital basis set, the situation may be more delicate. For these, the self-consistent cycle is usually performed over the single-particle density matrix, a quantity easily accessible from the Kohn–Sham orbitals. The same density matrix is typically used to re-start a DFT run, so that often the possibility to initialize a calculation from the real-space charge density is not directly available.

The work presented here is developed over the open-source Python package, PySCF,[44,45] which is used both for creating the training/test sets of the JLCDM and for running the MD. PySCF is an efficient numerical implementation of DFT (and also a variety of quantum chemistry methods) based on pseudopotentials and a Gaussian basis set. In the construction of the full DFT Fock matrix, among the charge-density-dependent terms, the exchange–correlation potential, $v_{\mathrm{XC}}(\mathbf{r})$, can be computed directly from $\rho(\mathbf{r})$ by using the Libxc library.[48] In contrast, the Hartree potential,

$$v_{\mathrm{H}}(\mathbf{r}) = \int \frac{\rho(\mathbf{r}')}{|\mathbf{r}-\mathbf{r}'|} d\mathbf{r}', \tag{2}$$

is not available. For this reason, we need to first develop an efficient numerical strategy for the calculation of $v_{\mathrm{H}}(\mathbf{r})$, a method that is presented next.

## B. Calculation of the matrix elements of $v_{\mathrm{H}}(\mathbf{r})$

The goal of this section is to present our numerical algorithm for the calculation of the Coulomb matrix, also known as the classical electron-repulsion matrix. The generic matrix element, computed between two basis functions, $\phi_\alpha(\mathbf{r})$ and $\phi_\beta(\mathbf{r})$, writes

$$\begin{aligned}\langle\phi_\alpha|v_{\mathrm{H}}|\phi_\beta\rangle &= \iint \frac{\rho(\mathbf{r}')}{|\mathbf{r}-\mathbf{r}'|}\phi_\alpha^*(\mathbf{r})\phi_\beta(\mathbf{r})d\mathbf{r}d\mathbf{r}' \\ &= \int \rho(\mathbf{r}')\left[\int \frac{\phi_\alpha^*(\mathbf{r})\phi_\beta(\mathbf{r})}{|\mathbf{r}-\mathbf{r}'|}d\mathbf{r}\right]d\mathbf{r}' \\ &= \int \rho(\mathbf{r}')\Omega_{\alpha\beta}(\mathbf{r}')d\mathbf{r}'. \end{aligned}\tag{3}$$

The second equality is used to highlight the structure of the equation. In fact, the term within the brackets does not depend on the charge density and it is just the real space representation of the matrix elements of the operator $\frac{1}{|\mathbf{r}-\mathbf{r}'|}$, which are here denoted as, $\Omega_{\alpha\beta}(\mathbf{r})$. Once this is computed, $\langle\phi_\alpha|v_{\mathrm{H}}|\phi_\beta\rangle$ can be obtained by direct integration with the ML charge density. Clearly, such a structure also suggests that the two integrations do not need to be carried out with the same grid. In the case of a representation based on Gaussian-type orbitals (GTOs), the calculation of $\Omega_{\alpha\beta}(\mathbf{r})$ is drastically simplified by the various analytic integrals available for Gaussian functions, as it will be demonstrated here. In fact, we will show that the calculation of $\Omega_{\alpha\beta}(\mathbf{r})$ is entirely analytical and no numerical integration needs to be performed.

Consider a generic GTO centered at the position $\mathbf{A}$ with angular momentum $l_a + m_a + n_a$, which is defined as

$$G_a = G_a(\mathbf{r},\alpha_a,\mathbf{A},l_a,m_a,n_a) = x_A^{l_a}y_A^{n_a}z_A^{m_a}e^{-\alpha_a r_A^2}, \tag{4}$$

where $\mathbf{r}_A = \mathbf{r}-\mathbf{A} = (x_A, y_A, z_A)$ and $r_A = \sqrt{x_A^2+y_A^2+z_A^2}$. The product of two GTOs, $G_1(\mathbf{r},\alpha_1,\mathbf{A},l_1,m_1,n_1)$ and $G_2(\mathbf{r},\alpha_2,\mathbf{B},l_2,m_2,n_2)$, centered, respectively, at $\mathbf{A}$ and $\mathbf{B}$ can be written as[49]

$$G_1G_2 = x_A^{l_1}x_B^{l_2}y_A^{m_1}y_B^{m_2}z_A^{n_1}z_B^{n_2}e^{-\alpha_1\alpha_2(\overline{\mathbf{AB}})^2/\gamma}e^{-\gamma r_P^2}, \tag{5}$$

where $\gamma = (\alpha_1+\alpha_2)$, $\gamma\mathbf{P} = \alpha_1\mathbf{A}+\alpha_2\mathbf{B}$, $\overline{\mathbf{AB}} = (\mathbf{A}-\mathbf{B})$, and $\mathbf{r}_P = \mathbf{r}-\mathbf{P}$.

For $s$-type orbitals ($l+n+m=0$), Eq. (5) is a simple Gaussian function centered at $\mathbf{P}$, while for non-zero angular momentum, the expression gets more involved because of the polynomial terms. In general, however, the use of Cartesian Gaussian functions allows one to write $G_1G_2$ as the product of three independent Cartesian components, e.g., $x_A^{l_1}x_B^{l_2}e^{-\gamma x_P^2}$, and a constant $e^{-\alpha_1\alpha_2(\overline{\mathbf{AB}})^2/\gamma}$. Furthermore, the product of the Cartesian polynomials can be written as the summation of various powers of the corresponding Cartesian components of the vector $\mathbf{r}_P$. For example, for the $x$ component, we write

$$x_A^{l_1}x_B^{l_2} = \sum_{k=0}^{l_1+l_2} x_P^k f_k^x(l_1,l_2,\overline{\mathbf{PA}}_x,\overline{\mathbf{PB}}_x), \tag{6}$$

with similar expressions holding for $y$ and $z$. In Eq. (6), $x_P$ is the $x$ component of the vector $\mathbf{r}_P$, while $\overline{\mathbf{PA}}_x$ and $\overline{\mathbf{PB}}_x$ are those of the vectors $\mathbf{P}-\mathbf{A}$ and $\mathbf{P}-\mathbf{B}$, respectively. Finally, the coefficients of the expansion read

$$f_k^x(l_1,l_2,\overline{\mathbf{PA}}_x,\overline{\mathbf{PB}}_x) = \sum_{i=0}^{l_1}\sum_{j=0}^{l_2}(\overline{\mathbf{PA}})_x^{(l_1-i)}\binom{l_1}{i}(\overline{\mathbf{PB}})_x^{(l_2-j)}\binom{l_2}{j}, \tag{7}$$

where, in taking the two sums, $k=(i+j)$ holds. A detailed derivation, taken from Ref. 49, can be found in Appendix A for completeness.

Since PySCF uses GTO functions as basis set, we can avail of the expression above to compute the matrix elements of the Coulomb matrix. In particular, we focus on computing $\Omega_{\alpha\beta}(\mathbf{r})$. For $\phi_\alpha = G_1$ and $\phi_\beta = G_2$, and considering that the GTOs are real, we have

$$\Omega_{12}(\mathbf{r}') = N_{12}K\sum_l\sum_m\sum_n f_l^x f_m^y f_n^z \int x_P^l y_P^m z_P^n \frac{e^{-\gamma r_P^2}}{|\mathbf{r}-\mathbf{r}'|}d\mathbf{r}, \tag{8}$$

where the sum over $l$, $m$, and $n$ run from zero to $(l_1+l_2)$, $(m_1+m_2)$ and $(n_1+n_2)$, respectively. Furthermore, we have introduced the shorthand definition $K = e^{-\alpha_1\alpha_2(\overline{\mathbf{AB}})^2/\gamma}$, while $N_{12}$ is the product of the normalization constants of $G_1$ and $G_2$ (a close expression for such product can be found in Ref. 50). In Eq. (8), the term $1/|\mathbf{r}-\mathbf{r}'|$ makes the integral not separable into Gaussian integrals of its Cartesian components. It is then convenient to rewrite such term through its Laplace transformation, namely, as

$$\frac{1}{|\mathbf{r}-\mathbf{r}'|} = \frac{1}{\sqrt{\pi}}\int_0^\infty e^{-s|\mathbf{r}-\mathbf{r}'|^2}s^{-\frac{1}{2}}ds. \tag{9}$$

This effectively takes $1/|\mathbf{r}-\mathbf{r}'|$ into a function looking as an $s$-type Gaussian with orbital exponent $s$ centered at $\mathbf{r}'$, leading to

$$\Omega_{12}(\mathbf{r}') = \frac{N_{12}K}{\sqrt{\pi}}\sum_l\sum_m\sum_n f_l^x f_m^y f_n^z I_{xyz}^{lmn}(\mathbf{r}'), \tag{10}$$

where we have defined

$$I_{xyz}^{lmn}(\mathbf{r}') = \int_0^\infty s^{-\frac{1}{2}}\left[\int_{-\infty}^{+\infty} x_P^l y_P^m z_P^n e^{-\gamma r_P^2}e^{-s|\mathbf{r}-\mathbf{r}'|^2}d\mathbf{r}\right]ds. \tag{11}$$

In order to integrate the expression for $I_{xyz}^{lmn}(\mathbf{r}')$, Eq. (11), we first apply the Gaussian product theorem for two exponents,

$$e^{-\gamma r_P^2}e^{-s|\mathbf{r}-\mathbf{r}'|^2} = e^{-\gamma|\mathbf{r}-\mathbf{P}|^2}e^{-s|\mathbf{r}-\mathbf{r}'|^2} = e^{-\frac{\gamma s|\mathbf{r}'-\mathbf{P}|^2}{(\gamma+s)}}e^{-(\gamma+s)|\mathbf{r}-\mathbf{D}|^2}, \tag{12}$$

where $\mathbf{D} = \frac{\gamma\mathbf{P}+s\mathbf{r}'}{\gamma+s}$. This allows us to perform the integration over $\mathbf{r}$ of the Gaussian integrals containing the Cartesian components of $\mathbf{r}-\mathbf{P}$, namely, $x_P^l$, $y_P^m$, and $z_P^n$. These integrals, in fact, are now separable. One can then perform a change of variable,

$$(x-P_x)^l = (x_D - \overline{\mathbf{PD}}_x)^l = \sum_{l'=0}^{l} (x_D)^{l'} \overline{\mathbf{PD}}_x^{(l-l')} \binom{l}{l'}, \tag{13}$$

where, as before, $x_D$ and $\overline{\mathbf{PD}}_x$ are the $x$ components of the vectors $\mathbf{r}-\mathbf{D}$ and $\mathbf{P}-\mathbf{D}$, respectively. Similar expression holds for both the $y$ and $z$ components. With this decomposition at hand, the integrals over the Cartesian components of $\mathbf{r}$ reduce to a sum of terms of the type,

$$\int_{-\infty}^{+\infty} (x-D_x)^{2j} e^{-(\gamma+s)(x-D_x)^2} dx = \frac{(2j-1)!!\sqrt{\pi}}{2^j(\gamma+s)^j\sqrt{\gamma+s}}. \tag{14}$$

Finally, by writing explicitly the expression for the powers of the $q$ component of the vector $\mathbf{P}-\mathbf{D}$, namely, $\overline{\mathbf{PD}}_q^n = \left[\frac{s}{\gamma+s}\right]^n (P_q - q)^n$, we obtain

$$I_{xyz}^{lmn}(\mathbf{r}') = \pi \sum_{l'=0}^{l/2} \sum_{m'=0}^{m/2} \sum_{n'=0}^{n/2} \binom{l}{2l'}\binom{m}{2m'}\binom{n}{2n'} \times \frac{(2l'-1)!!(2m'-1)!!(2n'-1)!!}{2^{l'+m'+n'}} \times (P_x - x')^{l-2l'}(P_y-y')^{m-2m'}(P_z-z')^{n-2n'} \times I'(\mathbf{r}'), \tag{15}$$

where

$$I'(\mathbf{r}') = \int_0^\infty \frac{s^{L-2L'} s^{-1/2} e^{-\frac{\gamma s|\mathbf{r}'-\mathbf{P}|^2}{(\gamma+s)}}}{(\gamma+s)^{L-L'}(\gamma+s)^{3/2}} ds = \sum_{h=0}^{L'} \frac{(-1)^h}{\gamma^{L-L'+h+1.5}} \times \binom{L'}{h} \times \frac{\Gamma[L-2L'+h+0.5]\Gamma_{\text{inc}}[(L-2L'+h+0.5), \gamma|\mathbf{r}'-\mathbf{P}|^2]}{|\mathbf{r}'-\mathbf{P}|^{2(L-2L'+h)+1}}, \tag{16}$$

with $L = l+m+n$ and $L' = l'+m'+n'$. In the equation above, $\Gamma$ and $\Gamma_{\text{inc}}$ are the gamma function and the incomplete gamma function, respectively, computed via the numerical recipe included in the SciPy library.[51] A more detailed derivation of $I'(\mathbf{r}')$ is provided in Appendix B.

Equations (15) and (16) fully define the matrix elements $\Omega_{12}(\mathbf{r})$, introduced in Eqs. (8) and (10). These are fully analytical and their evaluation has been implemented in a Python module, allowing multi-processing parallelization. The generic matrix element of the Coulomb matrix, Eq. (3), can then be computed by further integrating over $\mathbf{r}'$ the product of $\Omega_{12}(\mathbf{r}')$ with the charge density. This last integration is performed numerically over the PySCF real space grid by using the $N$-point numerical quadrature method. Since the Coulomb matrix is symmetric for real atomic orbitals, in practice only the matrix elements of the upper triangle are explicitly computed. Since all the other matrix elements of the Kohn–Sham Hamiltonian are already provided by PySCF, the access to the Coulomb matrix effectively enables us to start a PySCF calculation from a real-space density. An efficient way to compute the converged charge density from machine learning, the JLCDM, will be briefly discussed in Sec. II C.

### C. Jacobi–Legendre charge density model

In this sub-section, we introduce the JLCDM,[37,39] which is used here to predict the electron density over a real-space grid. The main idea of the model is to express the electron charge density at a grid point, $\mathbf{r}_g$, as a sum of many-body contributions, namely,

$$\rho(\mathbf{r}_g) = \rho(\mathbf{r}_g)^{(1)} + \rho(\mathbf{r}_g)^{(2)} + \rho(\mathbf{r}_g)^{(3)} + \cdots + \rho(\mathbf{r}_g)^{(n)}, \tag{17}$$

where $\rho(\mathbf{r}_g)^{(m)}$ is the $m$-body term, meaning that the charge density at $\mathbf{r}_g$ is determined by clusters of $m$ atoms. Such term writes

$$\rho(\mathbf{r}_g)^{(m)} = \sum_{i_1,i_2,\ldots,i_m}^{\text{unique}} \rho(\mathbf{r}_g)^{(m)}_{i_1,i_2,\ldots,i_m}, \tag{18}$$

where $\rho(\mathbf{r}_g)^{(m)}_{i_1,i_2,\ldots,i_m}$ is the contribution to the charge density at $\mathbf{r}_g$ originating from the atoms $i_1, i_2, \ldots, i_m$. The indices $i_1, i_2, \ldots, i_m$ run over all the atoms included within a sphere of radius $r_{\text{cut}}$ from $\mathbf{r}_g$, and the sum extends over all the unique combinations of the $m$ indices for distinct atoms.

The dependence of any $\rho(\mathbf{r}_g)^{(m)}_{i_1,i_2,\ldots,i_m}$ term over the atomic positions is then expressed by mean of Jacobi–Legendre polynomials. For instance, for the one- and two-body terms, those used here, we write

$$\rho(\mathbf{r}_g)_i^{(1)} = \sum_n^{n_{\max}} a_n^{Z_i} \tilde{P}_{nig}^{(\alpha,\beta)},$$
$$\rho(\mathbf{r}_g)^{(2)} = \sum_{n_1,n_2}^{n_{\max}} \sum_l^{l_{\max}} a_{n_1 n_2 l}^{Z_i Z_j} \tilde{P}_{n_1 ig}^{(\alpha,\beta)} \tilde{P}_{n_2 jg}^{(\alpha,\beta)} P_l^{ijg}. \tag{19}$$

The logic behind this representation is rather simple. The one-body term depends only on the distance between the grid point, $\mathbf{r}_g$, and the atom $i$ located at the position $\mathbf{r}_i$, $r_{ig} = |\mathbf{r}_i - \mathbf{r}_g|$. Such dependence is expanded over the "vanishing" Jacobi polynomials, $\tilde{P}_{nig}^{(\alpha,\beta)}$, defined as

$$\tilde{P}_{nig}^{(\alpha,\beta)} = \tilde{P}_n^{(\alpha,\beta)}\left(\cos\left(\pi\frac{r_{ig}-r_{\min}}{r_{\text{cut}}-r_{\min}}\right)\right), \tag{20}$$

$$\tilde{P}_n^{(\alpha,\beta)}(x) = \begin{cases} P_n^{(\alpha,\beta)}(x) - P_n^{(\alpha,\beta)}(-1) & \text{for } -1 \le x \le 1, \\ 0 & \text{for } \le -1. \end{cases} \tag{21}$$

Here, $P_n^{(\alpha,\beta)}(x)$ is the Jacobi polynomial of type $(\alpha,\beta)$ and order $n$, so that $\tilde{P}_n^{(\alpha,\beta)}(x)$ is constructed to vanish at the cutoff radius, $r_{\text{cut}}$. The expansion also includes a shift radius, $r_{\min}$, which is kept to zero throughout this work. The expansion coefficients $a_n^{Z_i}$ depends on the atomic species, $Z_i$, and the vector $(r_{\text{cut}}, r_{\min}, \alpha, \beta, n_{\max})$ contains the hyperparameters of the model.

The two-body contributions are constructed with the same logic. This time, a cluster comprising two atoms and the grid point defines a triangle, which is uniquely described by two distances, $r_{ig}$ and $r_{jg}$, and the angle subtended at $\mathbf{r}_g$. The expansion of the two-body terms is then made of products of two Jacobi polynomials, one for each distance, and a Legendre polynomial, $P_l^{ijg} = P_l(\hat{\mathbf{r}}_{ig} \cdot \hat{\mathbf{r}}_{jg})$, for the angle where $\hat{\mathbf{r}}_{kg} = (\mathbf{r}_k - \mathbf{r}_g)/r_{kg}$. Note that the Legendre polynomials are a special case of the Jacobi ones, obtained at $(\alpha,\beta) = (0,0)$. In addition, note that since one expects in the vicinity of an atom the one-body term to dominate the expansion, the vanishing Jacobi polynomials are replaced by the doubly vanishing ones in all the body-order terms higher than one. These read

$$\bar{P}_n^{(\alpha,\beta)}(x) = \tilde{P}_n^{(\alpha,\beta)}(x) - \frac{\tilde{P}_n^{(\alpha,\beta)}(1)}{\tilde{P}_1^{(\alpha,\beta)}(1)} \tilde{P}_1^{(\alpha,\beta)}(x) \quad \text{for} \quad n \geq 2. \tag{22}$$

Finally, the expansion can be continued to higher body-orders by recalling that a $m$-body cluster centered on the grind point is uniquely specified by $m$ distances and $m(m-1)/2$ angles, and hence, the corresponding expansions will be the product of $m$ Jacobi and $m(m-1)/2$ Legendre polynomials (see Ref. 27 for more details).

The final cluster expansion is linear in the number of clusters and hence JL polymomials, so that the expansion coefficients can be learned by simple regression. Furthermore, since the value of the density at a grid point depends only on the positions of the atoms within the cutoff sphere, but not on other grid points, the inference of the entire density can be trivially parallelized for numerical efficiency. Importantly, since the standard output of DFT is the charge density over a grid, a single DFT calculation in principle provides a huge number of training points. However, many of them are not relevant for determining the observables, for instance, those corresponding to regions of space far away from the atoms, and there is significant redundancy, namely, many grid points share extremely similar local atomic environments. For this reason, some care has to be taken when creating the training dataset. Past experience[37,39] suggests that a good strategy is to include in the training set grid points distributed according to the probability function $\exp\left[-(1/\rho(\mathbf{r}))^2/2\sigma^2\right]$, where the width $\sigma$ is treated as an additional hyperparameter. These points are usually complemented by others, randomly distributed across the simulation cell, with the relative fraction that can be optimized. Here, we use a 90:10 split, with 90% of the training point being randomly distributed.

## D. Computational details

The molecule of interest for this work is gas-phase uracil, $C_4H_4N_2O_2$, for which both experimental and DFT-predicted IR spectra are available.[52–54] All the electronic calculations used here, including those that generate the training and test sets, have been performed with DFT-driven AIMD, as implemented in the PySCF package.[44,45] We have chosen a 6-311++G Gaussian basis set,[55] which describes uracil with 152 orbitals, and the local density approximation to the exchange and correlation energy, as parameterized by Vosko, Wilk, and Nusair.[56] In all cases, the second-order mixing scheme, as implemented in PySCF,[57,58] has been utilized to achieve DFT convergence, while the molecular dynamics is driven by the Large-scale Atomic/Molecular Massively Parallel Simulator (LAMMPS) package.[59] This means that PySCF provides the atomic forces and the total energy, regardless of the method of choice (either full self-consistently or from the JLCDM), and then LAMMPS integrates the equations of motion and returns the updated atomic coordinates. For C, N, and O, the core electrons are described by a compact effective core potential[45] with a relativistic nodal boundary condition.

The training configurations have been extracted from 3.5 ps-long MD simulations integrated at a time step of $10^{-3}$ ps and initiated with different velocities. In particular, we have trained two models, respectively, from data taken at 100 and 300 K, with the training sets containing 46 uncorrelated molecular conformations for each temperature. A given model is then tested on molecular conformation taken from MD runs performed at the same temperature of the training, with the addition of a 200 K trajectory computed with the 300 K model. The test set is formed by 50 configurations randomly taken from a 20 ps-long simulation, with $10^{-3}$ ps time resolution. For these, we run fully converged static DFT calculations, where we compute both the real-space charge density and the single-particle density matrix. All AIMD runs are performed using the Nosé–Hoover thermostat[60] in the *NVT* ensemble.

For the converge parameters chosen, each static DFT calculation is associated to a 152 × 152 single-particle density matrix, while the real-space charge density is written over 152 320 points arranged over an atom-centered grid. As mentioned in Sec. II C, only a subset of these points is used for the training. In particular, here the parameter $\sigma$ is optimized to select 10 000 points only, corresponding to 6.5% of the total. We then include 12 additional points corresponding to the positions of the 12 nuclei of the atoms in the molecule. Thus, several batches of 10 012 points are supplied to the model during the training.

The set of hyperparameters defining the JLCDM is optimized against the mean absolute error (MAE), the maximum error, and the coefficient of determination, $R^2$, of the electron density grid-point values. This optimization is performed manually, by simply varying all the hyperparameters over a uniform grid of values. The optimal hyperparameters are reported in Table I, while the errors for each model are presented in Table II. Note that we provide error values for both the real-space charge density, the quantity computed directly by the JLCDM, and for the single-particle density matrix obtained by two diagonalization steps of the Hamiltonian defined by such real-space density (two steps are necessary due to our optimization scheme). As such, for the density matrix, we do not report the $R^2$ value. Overall, the final JLCDM is constructed over 6180 features (the coefficients of the cluster expansion), which are computed by simple Ridge regression.

**TABLE I.** JLCDM optimized hyperparameters for the uracil molecule, $C_4H_4N_2O_2$. We report the cutoff radius, $r_{cut}$, the polynomial order $n_{max}$ and $l_{max}$, the polynomial type $(\alpha,\beta)$ for all the body-orders (BO) considered. In all cases, we take $r_{min} = 0$. The total number of features of the JLCDM is 6180, while the hyperparameter governing the grid-point sampling is $\sigma = 90$ bohrs$^3$ ($a_0^3$). For the 1B order $l_{max}$ is not defined.

| BO | $r_{cut}$ (Å) | $n_{max}$ | $l_{max}$ | $\alpha$ | $\beta$ |
|---|---|---|---|---|---|
| 1B | 2.90 | 60 | ⋯ | 1.0 | 1.0 |
| 2B | 2.90 | 12 | 5 | 1.0 | 1.0 |

**TABLE II.** Table summarizing the errors of the two JLCDM's constructed, using data from AIMD, respectively, at 100 and 300 K. The errors are against the test set of 65 uracil conformers obtained from an AIMD at the reported temperature, $T$. At 200 K, we test the JLCDM generated with the 300 K training set. We report mean absolute error (MAE), maximum error, $\delta_{\max}$, root-mean-square error (RMSE), and coefficient of determination $R^2$. We show data for the real-space charge density and for the density matrix (DM) obtained after one SCF step following the initial diagonalization of the Kohn–Sham Hamiltonian. The units for the real-space charge density are electron/bohr$^3$ $(e/a_0^3)$ and those for the density matrix are atomic units (a.u.).

| | $T$ (K) | MAE | $\delta_{\max}$ | RSME | $R^2$ |
|---|---|---|---|---|---|
| $\rho(\mathbf{r})$ | 100 | $1.10 \times 10^{-3}$ | 0.0835 | $2.79 \times 10^{-3}$ | 0.9997 |
| $\rho(\mathbf{r})$ | 200 | $1.50 \times 10^{-3}$ | 0.1077 | $3.70 \times 10^{-3}$ | 0.9995 |
| $\rho(\mathbf{r})$ | 300 | $1.60 \times 10^{-3}$ | 0.1012 | $3.89 \times 10^{-3}$ | 0.9994 |
| DM | 100 | $5.75 \times 10^{-5}$ | 0.1022 | $4.27 \times 10^{-4}$ | ... |
| DM | 200 | $2.10 \times 10^{-4}$ | 0.4933 | $2.05 \times 10^{-3}$ | ... |
| DM | 300 | $3.70 \times 10^{-4}$ | 0.5274 | $2.34 \times 10^{-3}$ | ... |

## III. RESULT AND DISCUSSIONS

Our JLCDM is evaluated in three different steps. First, we assess the accuracy of the computed real-space charge density and that of the one-particle density matrix constructed from a single self-consistent-field (SCF) step initiated at the Kohn–Sham Hamiltonian built from the ML charge density (two diagonalization steps in total). We will call the latter quantity the JL density matrix (JLDM). This gives us an estimate of the accuracy of the electronic quantities predicted by the JLCDM. Then, we compare the atomic forces computed from the JLDM against those obtained from performing a full self-consistent cycle for a molecule of the same geometry. This test allows us to assess how the JLCDM charge density can drive AIMD. Finally, we look at the thermal dependence of electronic observables. In particular, we compute the IR spectrum by using all JLCDM-derived quantities, namely, the forces and the electrical dipole obtained from the JLDM.

### A. JLCDM charge density and JLDM

In Fig. 1, we present the parity plots for both the JLCDM charge density [panels (a), (c), and (e)] and the corresponding JLDM [panels (b), (d), and (f)]. These are computed over 65 configurations extracted from AIMD trajectories run at 100 K [panels (a) and (d)], 200 K [panels (b) and (e)], and 300 K [panels (c) and (f)]. Each JLCDM is tested at its corresponding training temperature, with the exception of the 200 K test, which uses the JLCDM constructed at 300 K. Data are presented on a log–log scale to identify better deviations across the entire range of values. Considering first the charge density, it is clear that an excellent agreement is in general obtained for densities larger than $10^{-2}$ $e/a_0^3$ (electrons/bohr$^3$). On the scale of the graphs, it is difficult to distinguish among the different models, which appear to perform equally well. A more precise evaluation can be obtained by looking at the computed metrics, namely, the mean absolute error (MAE), maximum error, $\delta_{\max}$, and root-mean-square error (RMSE), which are reported in Table II. These show a minor degradation in performance as the testing temperature is increased, an error associated to a larger diversity in the geometrical configurations explored at high temperature. The changes, however, are extremely minor and we can conclude that the two models perform approximately equal. Note that these errors are about one order of magnitude larger than what were previously found for JLCDMs trained over the valence charge of pseudopotential plane-wave calculations.[37,39] We attribute such difference to details in the JLCDMs and the datasets. The current model can certainly be improved by enlarging the cluster expansion to higher-body order terms, a task that was not executed here in order to keep the model slim. This, in fact, must have an inference speed competitive with that of a standard SCF procedure. As we will show later in this section, the current model is already of sufficient quality to compute the IR spectrum without performing SCF convergence, giving an overall workflow computationally competitive to fully converged AIMD.

Looking now at the density matrix, panels (b), (d), and (f), also in this case the agreement between the fully converged DFT values ($\mathrm{DM}^{\mathrm{DFT}}$) and those obtained with a single SCF step from the Hamiltonian constructed over the ML charge density is pretty satisfactory. The errors reported here are similar to those obtained previously with a fully connected neural network predicting directly the density matrix,[43] a result suggesting that the quality of our model is sufficient to drive the MD. Further evidence is provided by the error over the total energy computed from the JLDM. In this case, we find average absolute deviations from the fully DFT-converged values of $2.67 \times 10^{-5}$, $4.58 \times 10^{-4}$, and $1.2 \times 10^{-3}$ hartree, for the data at 100, 200, and 300 K, respectively. Note that the total energy of uracil is of the order of $-76.5$ hartree, so that the errors are around $3.5 \times 10^{-5}$%, $6.0 \times 10^{-4}$%, and $1.7 \times 10^{-3}$%, respectively. Finally, we have tested the distance of JLDM from the fully converged one. This is evaluated as

$$\delta\mathrm{DM} = \frac{\sum_{ij} |\mathrm{DM}_{ij}^{\mathrm{DFT}} - \mathrm{DM}_{ij}^{\mathrm{ML}}|}{\sum_{ij} |(\mathrm{DM}_{ij}^{\mathrm{DFT}} + \mathrm{DM}_{ij}^{\mathrm{ML}})/2|} \times 100,$$

where $\mathrm{DM}_{ij}^{\mathrm{ML}}$ is the $(ij)$ matrix element of the JLDM, while $\mathrm{DM}_{ij}^{\mathrm{DFT}}$ is that of the fully DFT-converged DM for a molecule having the same geometry. This error is computed to be 0.19%, 0.73%, and 1.36% for molecules, respectively, at 100, 200, and 300 K.

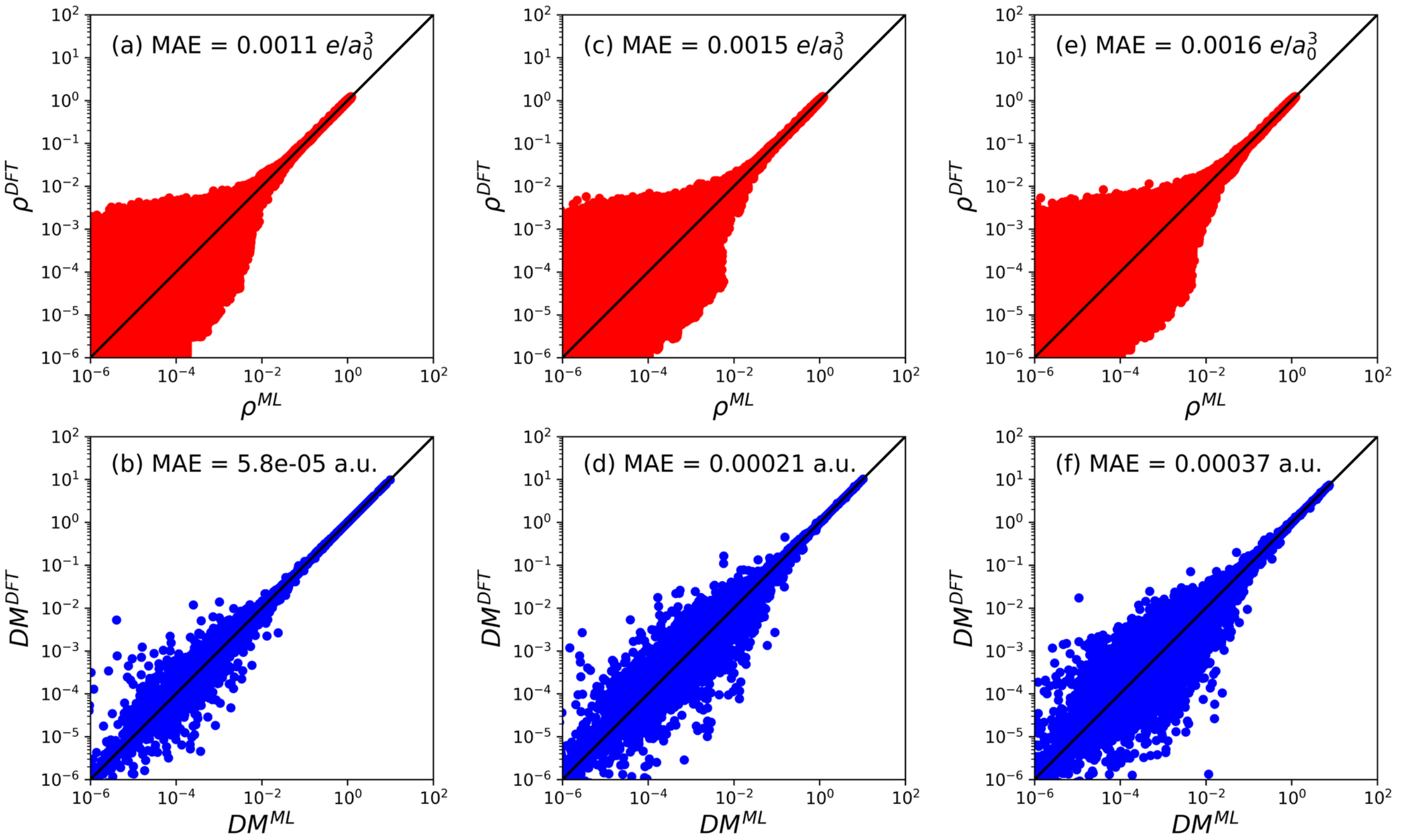

**FIG. 1.** Parity plots for the real-space charge density [panels (a), (c), and (e)] and the associated JLDM [panels (b), (d), and (f)] for the uracil molecule, with geometries taken from AIMD trajectories at three different temperatures: 100 K [panels (a) and (d)], 200 K [panels (b) and (e)], and 300 K [panels (c) and (f)]. Since the graphs are in log-scale, for the JLDM we plot the absolute value of the matrix elements. In all cases, the JLCDM values are along the *x* axis, while the fully converged DFT ones are on the *y* one.

## B. Forces evaluated from the JLCDM charge density

Next, we move to evaluate the accuracy of the atomic forces obtained from a single SCF step of the Kohn–Sham Hamiltonian constructed from the JLCDM. In Fig. 2, we show parity plots of the $\alpha = x, y, z$ components of the forces computed over the 100 and 200 K test set, while those for the 300 K case are presented in the supplementary material. The figure also reports histograms of the forces distributions, which become broader as the temperature is increased, as expected.

We find an excellent agreement between the ML-computed force components, $F_\alpha^{\mathrm{ML}}$, and those obtained by converging DFT at the same geometry, $F_\alpha^{\mathrm{DFT}}$. The MAEs for the *x*, *y*, and *z* components are, respectively, 2.8, 3.1, and 1.9 meV/Å at 100 K, then grow to 9.1, 10.2, and 12.0 meV/Å at 200 K, and finally to 24.8, 28.2, and 16.4 meV/Å at 300 K. A similar behavior is found for the maximum deviation, which goes from 91.9, 69, 134.8 meV/Å (*x*, *y*, and *z* components) at 100 K, to 418.5, 704.3, 801.6 meV/Å at 200 K, to 1127.69, 1016.8, 1093.3 meV/Å at 300 K. In general, the model trained at 100 K performs better than that at 300 K, with a difference in the MAE of approximately an order of magnitude. This is somehow expected since the two models have an identical number of parameters, but the model trained at 300 K explores a larger configuration space, given by the stronger thermal agitation of the molecules. In addition, as expected, the 300 K model, when tested over molecules with configurations taken from a 200 K AIMD simulation, performs better than when tested on configurations generated at 300 K. This is consistent with having a narrower distribution of molecular geometries at the lower temperature. Note that the maximum error, although increases with temperature, shows some anomalies when broken down to the individual components of the forces. This originates from the relatively small test set pool, only 65 configurations, and from the fact that the test molecules can be slightly rotated with respect to each other (as driven by the MD), so that the individual components may have some level of mixing.

Importantly, the errors found on the forces obtained from the JLCDM are significantly lower than those that we have previously computed with a neural network predicting the single-particle density matrix.[43] For instance, the forces on $S_2O$ for molecular geometries corresponding to an MD run at 150 K were found to have a MAE of about 100 meV/Å, a value that could be approximately halved by rescaling the total electron count. Note that the $S_2O$ molecule is simpler than uracil, studied here, and also that the average force components for that case were in the ±0.25 meV/Å range against ±1 meV/Å of the present situation. We attribute the better performance of the JLCDM to two main reasons. In the case of the JLCDM, the single-particle density matrix that is used to compute the forces is obtained by diagonalizing the Kohn–Sham Hamiltonian. This ensures the density matrix is both idempotent and integrates to an integer number of electrons, two features that are not necessarily met by evaluating the matrix elements of the density matrix via a neural network. The total number of electrons can be imposed by simple re-scaling, but the idempotency is a non-linear conditions, difficult to enforce. As such, the current method appears to have a further advantage over the direct evaluation of the density matrix. Finally, note that the forces computed in Ref. 43 were already accurate enough to drive MD, resulting in trajectories with a correct bond distribution. This suggests that the same should be true for the forces generated here from the JLCDM, an aspect that will be investigated next.

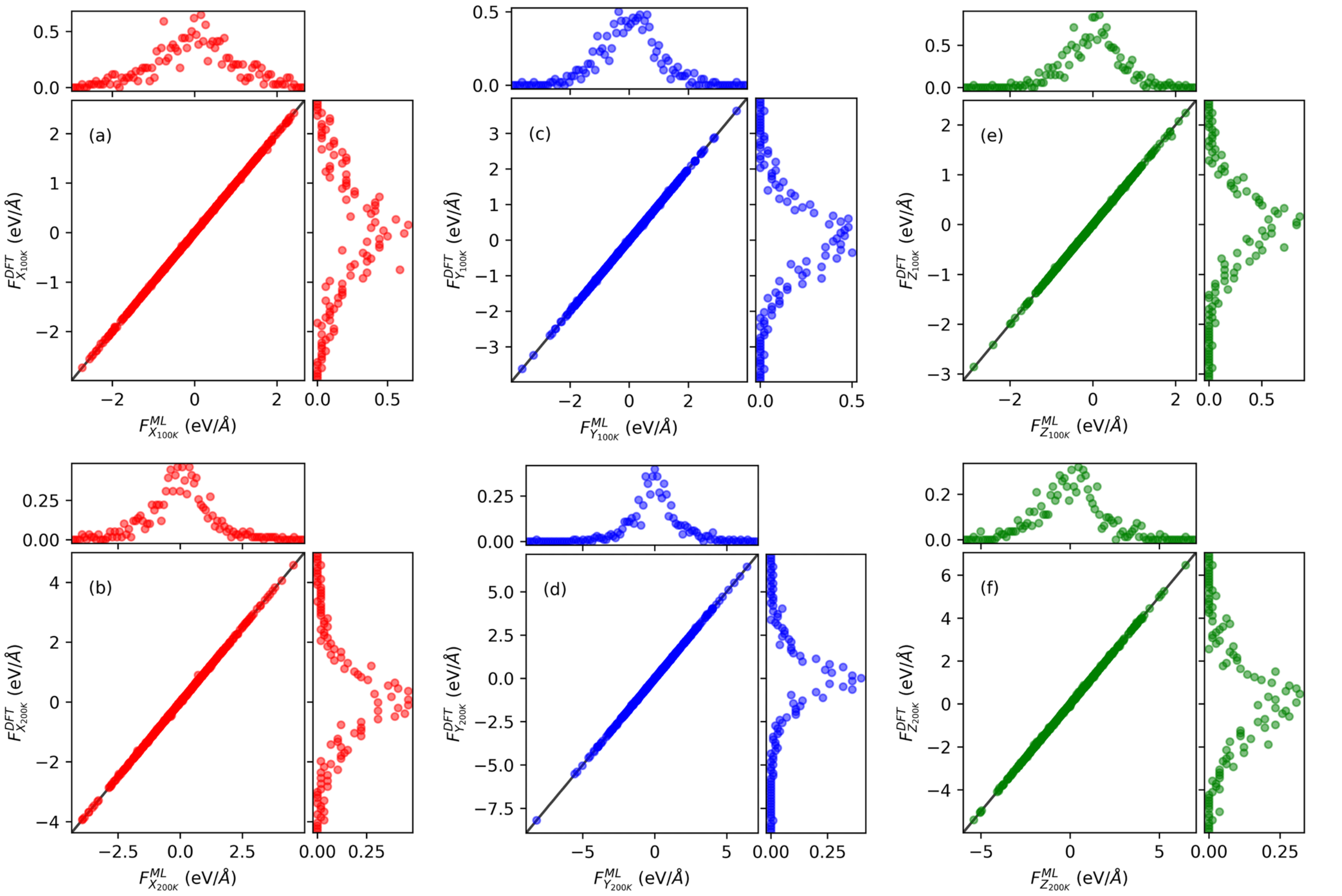

**FIG. 2.** Parity plots for the various components of the atomic forces. Here, $F^{\mathrm{ML}}_{\alpha}$ is the $\alpha$ component of the forces obtained after one SCF step starting from the Kohn–Sham Hamiltonian constructed with the charge density predicted by the JLCDM, while $F^{\mathrm{DFT}}_{\alpha}$ is the DFT-converged one. Panels (a), (c), and (e) are for the *x*, *y*, and *z* components of the forces computed over the 100 K dataset, while panels (b), (d), and (f) are for the 200 K one. The parity plots also show the forces histograms. Data for the 300 K configurations are presented in the supplementary material.

## C. JLCDM evaluation of the IR spectrum

We now present a possible ML workflow for computing a temperature-dependent electronic quantity. As discussed in Sec. I, here we consider the case of the IR spectrum, which can be obtained from the Fourier transform of the autocorrelation function of the dipole moment over an MD trajectory,[61]

$$\tilde{\mu}(\omega) = \int_{-\infty}^{+\infty} \langle \boldsymbol{\mu}(t) \cdot \boldsymbol{\mu}(t+\tau) \rangle e^{-i\omega\tau} \mathrm{d}\tau, \tag{23}$$

where $\boldsymbol{\mu}(t)$ is the dipole moment of the molecule at time $t$ and $\tau$ is the delay time, which is here taken as the MD time step, $\tau = 10^{-3}$ ps. The relative intensity of the IR spectrum can then be computed as

$$I(\omega) \propto \omega |\tilde{\mu}(\omega)|^2 \left[ 1.0 - e^{\frac{-\hbar\omega}{K_B T}} \right], \tag{24}$$

where $K_{\mathrm{B}}$ is the Boltzmann's constant. Here, we discuss simulations at both 100 and 200 K, for which we compare the IR spectra obtained with fully converged AIMD and with the JLCDM. We have also performed simulations at 300 K, which turned out to be rather unstable. The main reasons behind such instability will be discussed here, while more details are presented in the supplementary material.

The computed IR spectra are presented in Fig. 3, where panels (a) and (b) are taken at 100 K, while panels (c) and (d) are for 200 K. In both cases, the spectra are obtained from MD simulations 20 ps-long, with the top panels corresponding to the predictions of the JLCDM, while the bottom two are for fully converged AIMD. Note that in the case of the JLCDM, the ML model drives the MD and also provides the value of the dipole moment, so that there is no self-consistency at any level. In general, we find a good agreement between the JLCDM- and the AIMD-calculated spectra at both the temperatures investigated. The JLCDM spectrum is slightly more noisy than the AIMD one across the entire spectral region investigated, as a result of the noise in the atomic forces. This is particularly evident in the frequency region 2000–3500 $cm^{-1}$, where the JLCDM spectrum presents a small background signal, absent in the AIMD one. Such noise grows with temperature, as a reflection of the increased error in the forces, but the JLCDM still has sufficient accuracy to well describe the peak broadening and possible drift.

When comparing the spectra at a more quantitative level, we observe that the maximum relative intensity at 100 K is found at 1736.37 $cm^{-1}$ by AIMD and at 1733.04 $cm^{-1}$ by JLCDM. The same peaks blueshift at 1718.22 and 1716.49 $cm^{-1}$ at 200 K, respectively. This spectral region corresponds to the C2O and C4O stretching modes,[52,53] although an exact attribution is complicated and beyond the scope of this work. Previous DFT studies position[54] the maximum-intensity peak in between 1720 and 1813 $cm^{-1}$, depending on the choice of exchange–correlation functional, consistent with the results presented here. The agreement between the JLCDM and the AIMD spectra is also notable away from the high-intensity region, namely, around 500 $cm^{-1}$, where one finds modes related to the bending of the aromatic ring, in between 1000 and 1500 $cm^{-1}$, corresponding to various local bond stretching modes, and just above 3500 $cm^{-1}$, where the N1H mode is positioned. This last part of the spectrum becomes significantly more noisy when computed with the JLCDM at 200 K and the peak is barely visible from the background. Some noise can be possibly eliminated by performing longer MD trajectories, but not completely since the error in the forces discussed before.

Note that, at variance with experiments, our spectra do not display intensity in the interval 600–1000 $cm^{-1}$, with the exception of a small peak in the AIMD spectrum just below 800 $cm^{-1}$. This discrepancy is attributed to two main factors. On the one hand, the modes in the interval 600–1000 $cm^{-1}$ window have all low intensity, so that they may require much longer MD simulations to emerge. On the other hand, they appear to be sensitive to the fine detail of the total energy and depend strongly on the choice of functional and pseudopotentials. Finally, we wish to mention that we have also computed the IR spectrum at 300 K. This, however, turned out to be extremely noisy. The main issue found is that for some geometries, the JLCDM predicts an inversion between the highest-occupied molecular orbital (HOMO) and the lowest-unoccupied molecular orbital (LUMO). As a result, the molecule dipole moment jumps abruptly, leading to a high level of noise in the Fourier transform.

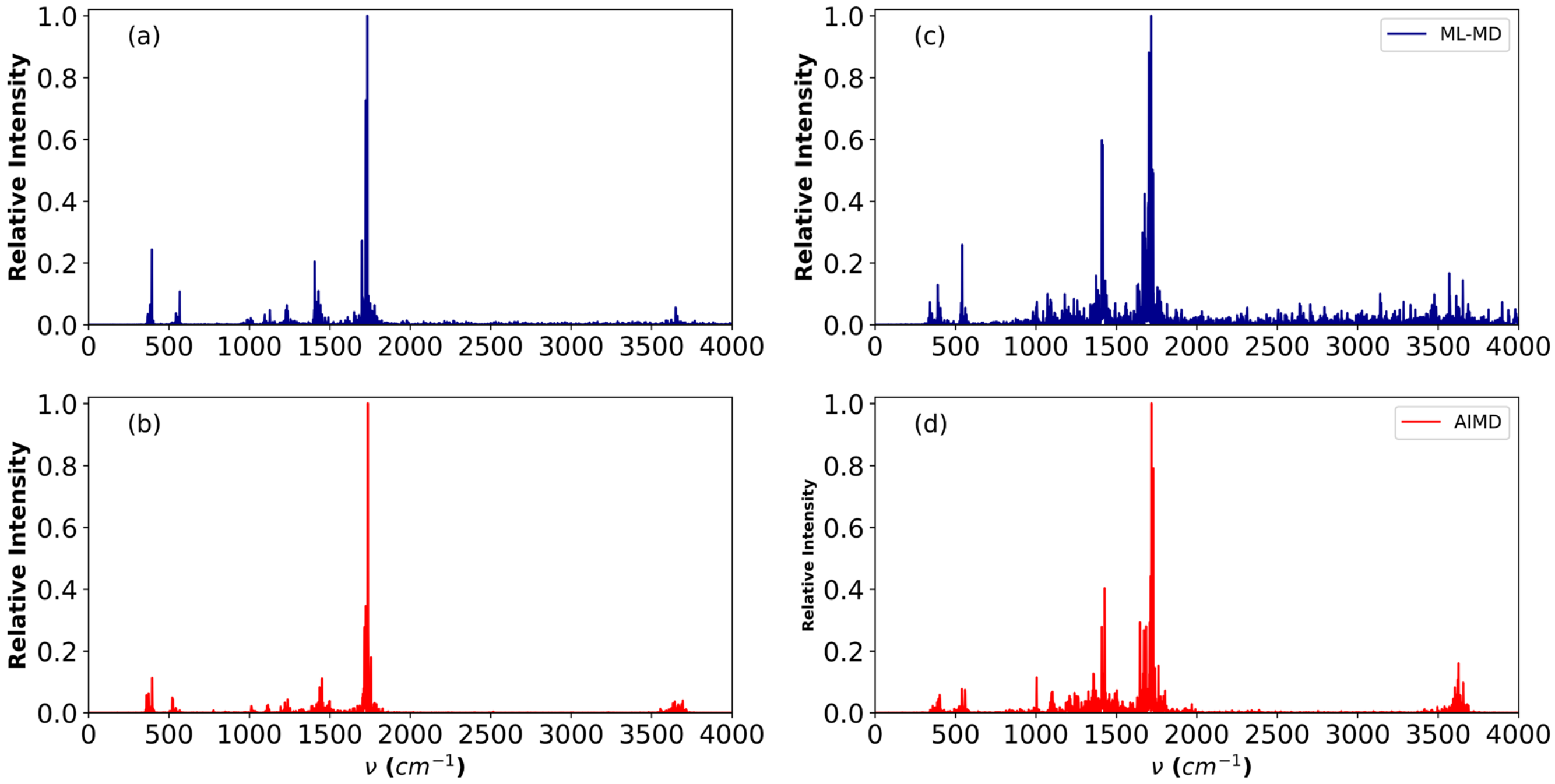

**FIG. 3.** IR spectra obtained from MD trajectories at 100 K [panels (a) and (b)] and 200 K [panels (c) and (d)]. The two top panels are for the JLCDM, while the bottom two correspond to fully converged AIMD. The figure displays the relative spectral intensity in the interval [0, 1].

More details about this failure are discussed in the supplementary material.

Finally, we wish to close this section by analyzing in detail the origin of the differences between the JLCDM- and the AIMD-predicted spectra. These boil down to two main causes: (i) the error in the determination of the molecular structure, arising from the inaccuracy in the calculation of the forces, and (ii) the error in the estimate of the dipole moment. These two sources of error are difficult to disentangle since there is a strong correlation between molecular geometry and dipole moment. For this reason, in Fig. 4 we show the parity plots for the Cartesian components of the dipole moment computed at different levels of theory. In particular, in the upper panels, we show parity between the absolute value of the JLCDM-computed dipole at the JLCDM-computed geometry, $|\mu_\alpha|^{\mathrm{ML}}_{R_{\mathrm{ML}}}$, with the same quantity evaluated at the fully converged DFT level, $|\mu_\alpha|^{\mathrm{DFT}}_{R_{\mathrm{ML}}}$. In contrast, the lowest panels show parity between the same quantities, but this time evaluated at the geometries obtained from AIMD, $|\mu_\alpha|^{\mathrm{ML}}_{R_{\mathrm{DFT}}}$, and $|\mu_\alpha|^{\mathrm{DFT}}_{R_{\mathrm{DFT}}}$. In both cases, the test is performed over 1000 uncorrelated configurations taken from a 20 ps-long MD trajectory in the interval 5–6.8 ps. Similar results at 200 K are presented in the supplementary material.

From the figure, it clearly appears that the JLCDM evaluation of the dipole moment is rather accurate, with no notable difference depending on whether the geometries were taken from the AIMD or the JLCDM trajectory. In general, no component has a MAE larger than 0.01 D and, considering that the total dipole moment is of the order of 4.8 D, this corresponds to an inaccuracy of only 0.2%. Similarly, the maximum error found is around 0.099 D, which corresponds to only 2%. The simulations at 200 K display only a very marginal increase in the MAE and in the maximum error (see the supplementary material), a fact that leads us to conclude that that difference in the spectra has to be attributed to the slightly different geometries encountered along the MD trajectories driven with different methods.

### D. JLCDM evaluation of other structural and electronic observables

Finally, we investigate the thermal distributions of a few electronic observables, namely, the total DFT energy, the HOMO–LUMO bandgap (computed from the Kohn–Sham energies), and the amplitude of the dipole moment. In this case, we take the average over the entire MD trajectories, either driven by DFT or by the JLCDM. The results for the 100 K simulations are presented in Fig. 5, while those at 200 K are included in the supplementary material.

In general, we find that the average values of each quantity are rather similar, regardless of how they are calculated. For instance, at 100 K, the average total energy is −76.517 hartree when computed from AIMD, while the JLCDM returns −76.516 hartree, with a difference of about 0.0013%. Similarly, we have average HOMO–LUMO gaps of 0.134 and 0.135 hartree, and dipole moments of 4.8695 and 4.8519 D, respectively, from the AIMD and the JLCDM. Such small differences persist when the temperature is increased to 200 K. More interestingly, we find that the distributions computed from the JLCDM-driven MD are systematically broader than those taken from AIMD, suggesting a larger diversity in the molecular conformations explored. This supports the previous observation that the main source of error in the JLCDM evaluation of the electronic quantities seem to arise from the molecular structures explored during the MD. It is important to note that, although generally small differences in the distributions may not present a particular problem, they might in the case of the HOMO–LUMO gap. In fact, we find that as the temperature increases (see 200 K data in the supplementary material), the lower end of the bandgap distribution is significantly reduced. This reduction becomes extreme at 300 K, where we encounter conformations presenting a HOMO–LUMO inversion, with a consequent discontinuous jump in all the electronic observables. This problem is at the origin of the failure of the JLCDM to generate an IR spectrum at 300 K.

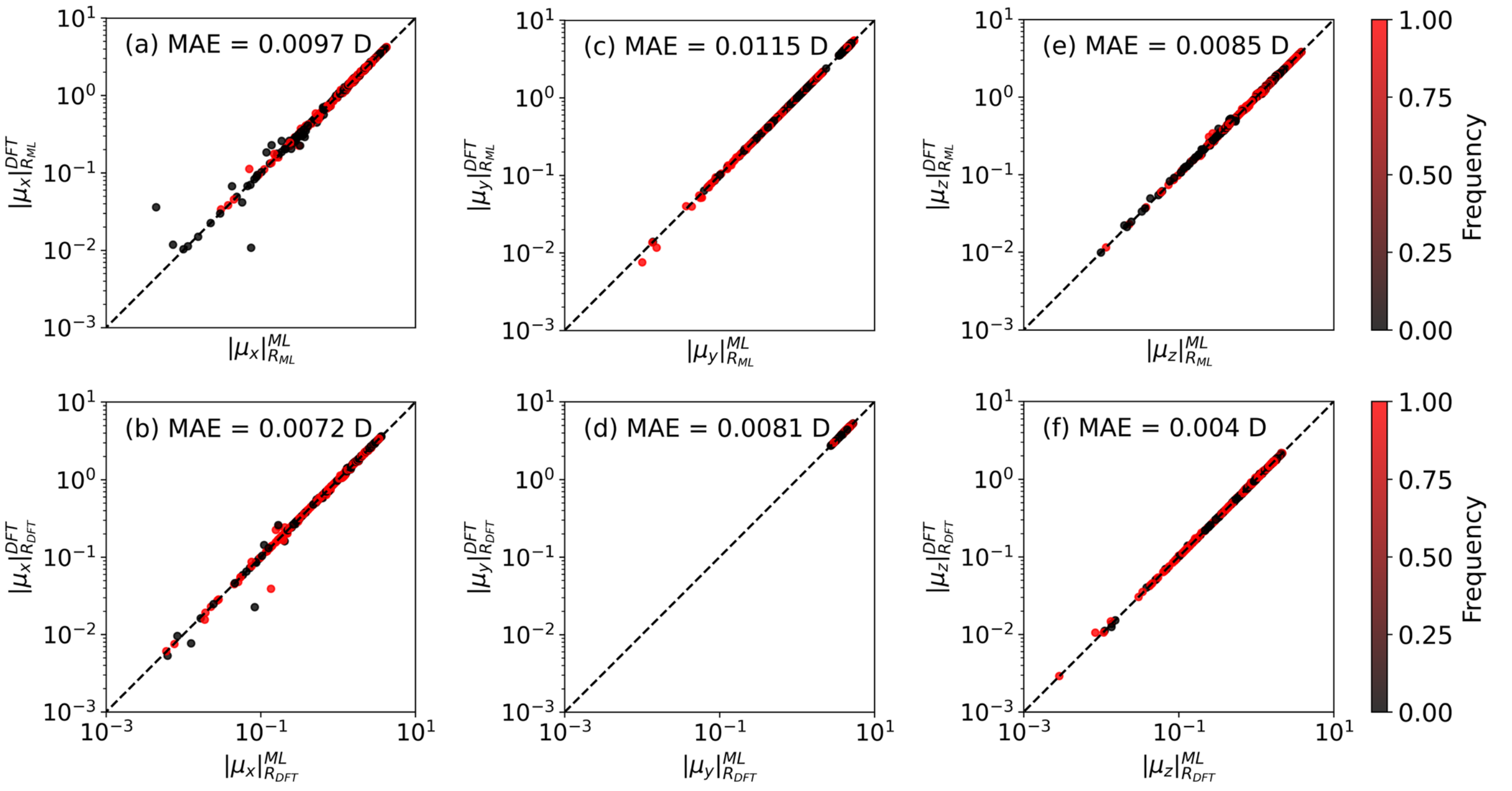

**FIG. 4.** Parity plots for the $\alpha = x, y, z$ components of dipole moments computed at 100 K. Here, we show results from 1000 uracil uncorrelated structures taken from a 20 ps-long MD trajectory. Here, $|\mu_\alpha|^{\gamma}_{R_\beta}$ is the absolute value of the $\alpha$ component of the dipole moment computed with the method $\gamma$ ($\gamma$ = DFT or ML) at the geometry driven by MD performed with $\beta$ ($\beta$ = DFT or ML). In all subplots, the units are Debye (D) and the scale is logarithmic (hence, we plot the absolute value of the given component). The color encodes the density of a given value, while the MAE is reported as a legend.

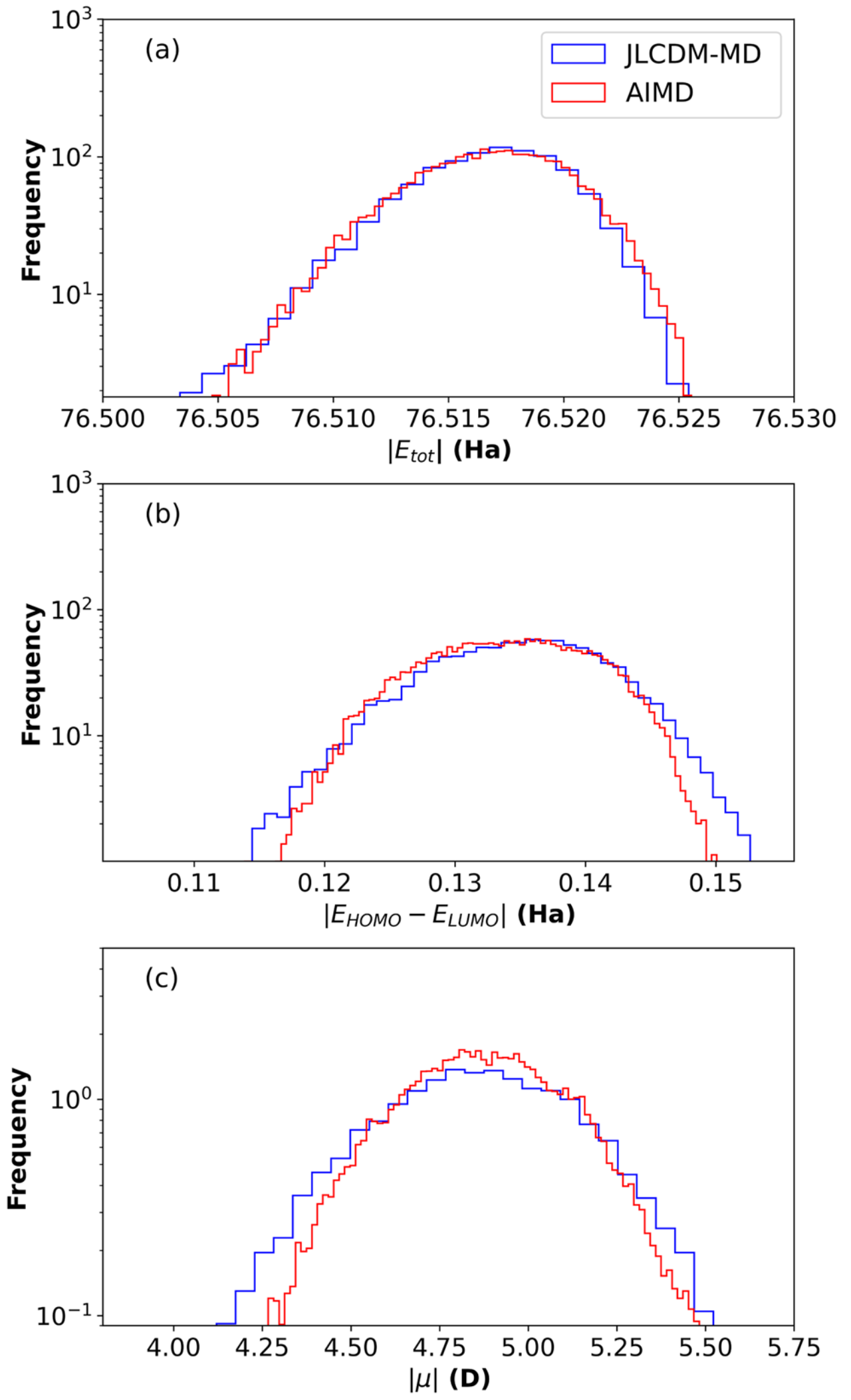

**FIG. 5.** Histogram of the thermal distribution at 100 K of (a) the total DFT energy, $E_{tot}$, (b) the HOMO–LUMO Kohn–Sham bandgap, $|E_{HOMO} - E_{LUMO}|$, and (c) the amplitude of the dipole moment, $|\boldsymbol{\mu}|$. These have been obtained over the entire trajectories, generated either with AIMD or from the MD driven by the JLCDM.

We would like to close the section with a brief discussion on the possible ways to improve our model. As with ML force fields, the JLCDM can also be systematically improved by extending the training set and by enlarging the model. The first task effectively consists in running further DFT calculations for additional structures, while the second will require including high body-order contributions to the cluster expansion. Both tasks require a systematic study since there are multi alternative possibilities, while the final goal remains that to generate a compact model trained on a limited number of examples. This systematic study is outside the scope of the present paper. Ideally, the optimization of the cluster expansion and the construction of an improved training set should be performed together in an active learning strategy.

In general, the inclusion of high body-order terms is not desirable and should be avoided if possible. This is because the inference time of the model depends on its complexity and it increases with the order of the terms considered. The enlargement of the training set, instead, does not add further complexity, but requires a clever choice of the new structures. These, in fact, need to be as representative as possible of the charge densities associated to the structures that one will encounter in a MD simulation. We envision two possible strategies. On the one hand, one can systematically evaluate the uncertainty on predictions for molecular configurations not sampled by the training set. This allows one to have a simple estimator of the diversity of the new configurations to select for the enlargement of the training set. Such a strategy has been applied successfully to ML force fields,[62] but never to a JLCDM. On the other hand, we can generate a training set from MD simulations run at different temperatures and construct configuration-selection algorithms that distribute the conformations according to their structural diversity. In addition, in this case, the approach has been demonstrated as successful for ML force fields running over an extended temperature range,[63] but never for a JLCDM.

## IV. CONCLUSION

We have here presented a novel machine-learning workflow for computing thermal-averaged electronic observables and applied it to the calculation of the IR spectrum of the uracil molecule in the gas phase. The main idea is to use the Jacobi–Legendre charge-density model to predict the charge density of a converged DFT calculation. This is then used to construct the Kohn–Sham Hamiltonian, which can return all the needed observable with a single diagonalization step. The scheme has been implemented within the DFT numerical implementation of PySCF, for which we have constructed a plugin to compute the Hartree potential. In a nutshell, our proposed workflow provides all the information of an AIMD, without the need to perform fully converged DFT calculations. As such, it has a number of advantages. First, at variance with force fields, we do not need two (or more) different models, one for energy and forces and others for the electronic observables since the same JLCDM both drives the MD and provides an estimate of the electronic quantities. Furthermore, the charge density appears easy to learn, with a single DFT calculation providing a large number of training points. In addition, when the density is computed over a real-space grid, it is sufficient for the model to be rotational invariant and no complex equivariance is needed. Finally, our approach seems much more effective and accurate than learning directly the single-particle density matrix. This is because it avoids problems related to the violation of the density-matrix idempotency and electron number conservation.

The proposed workflow was then used to predict the temperature-dependent IR spectrum of the uracil molecule in the gas phase. In general, we find a good agreement between the spectra computed with the JLCDM and those obtained with AIMD, in particular at low temperature. In that case, the positions of the high-intensity peaks are extremely close with errors on the energy below 1%. As the temperature is increased, the JLCDM-derived spectrum remains systematically more noisy than the AIMD one, when computed over trajectories of identical duration. This feature is associated to a larger range of molecular conformations explored by the JLCDM-driven MD, over those of the AIMD, a fact confirmed by the high accuracy in the determination of a dipole moments for fixed geometries. The situation deteriorates at 300 K, for which the trained JLCDM produces an extremely noisy spectrum. This is mainly due to molecular geometries where there is an inversion of the positions of the HOMO and LUMO. Since these orbitals have different symmetry, such an inversion results in a discontinuous jump in the dipole moment and the consequent noise in the autocorrelation function. There are several possible ways to fix the problem. In general, one can construct a more accurate JLCDM by including in the training set some molecules with conformations presenting a very small HOMO–LUMO gap, namely, at the boundary of the mentioned energy-level inversion. Alternatively, one can include conformations taken at a temperature significantly higher than that the model will be used for. A second possibility is to build a simple classifier that identifies critical molecular conformations. These can then be slightly modified to achieve a stable MD, or one can construct a second JLCDM specifically addressing these critical conformations.

The method introduced here has been applied to calculations performed at the local-density-approximation level using pseudopotentials, so that one may wonder about the generality of the proposed scheme. In principle, the JLCDM formalism is general and can predict any density written over a real space grid. This means that it can be used with any exchange–correlation functional as long as it is not orbital-dependent since the JLCDM does not provide the Kohn–Sham orbitals. In practice, however, one should expect some additional limitations related to the model accuracy and the training strategy. First, functionals using the density gradient or higher derivatives may require a more accurate charge density since derivatives are generally more numerically unstable than the density. Possible mitigation strategies may include interpolation schemes that smooth out the JLCDM-predicted real-space density. Then, we expect that fitting an all-electron density may be more demanding than fitting the valence density only. This is associated with the all-electron density spanning a much larger range of values than the valence one. Again, one can design mitigation strategies, for instance, by fitting only the valence component of an all-electron density, while simply adding the core part.

Finally, we wish to comment on the numerical efficiency of our workflow. At present, for the uracil molecule and our numerical implementation, the JLCDM strategy is not faster than conventional AIMD. In fact, when run on a single 32-CPU node, an AIMD step takes about 60 s, while the JLCDM is at about 200 (120 for the JLCDM inference and 80 for the construction of the Hamiltonian). This, however, is not a fair comparison since our strategy has not been optimized for speed yet. In general, the JLCDM is a linear model, so that its inference is numerically cheap. Moreover, since each grid point is independent, it can be trivially parallelized with 100% efficient scaling. Further efficiency can be obtained by implementing interpolation strategies, which will eliminate the need for running the JLCDM for every grid point. In particular, one can obtain drastic saving by minimizing the inference on grid points away from the atoms. Most importantly, our workflow becomes increasingly more competitive as (1) the complexity of the DFT functional increases and (2) the molecules get larger. These, in fact, are two situations where the numerical overheads of a fully converged self-consistent procedure drastically increase, while those of the JLCDM remain roughly unchanged. Finally, we wish to note that our method still remains grounded in DFT, so that it will unlikely compete in speed with semi-empirical approaches based on ML models only.[64] We believe that these are complementary approaches, whose benefits and drawbacks depend on the system at hand and the level of accuracy required.

In summary, we have here presented a ML workflow for the computation of the temperature-dependent IR spectrum of molecules. This is alternative to force fields and adds a novel weapon to the ML arsenal of numerical tools available for materials design.

## SUPPLEMENTARY MATERIAL

The supplementary material file contains (1) the forces evaluated from the JLCDM charge density at 300 K; (2) the parity plot for the dipole moment at 200 K; (3) the thermal distributions of various observables at 200 K; and (4) the IR spectrum computed at 300 K with an analysis of the error.

## ACKNOWLEDGMENTS

This work was supported by the Government of India, NOS Award (No. K-11015/65/2020-SCD-V/NOS). U.P. acknowledges the Qatar National Research Fund (Grant No. NPRP12S-0209-190063) for providing financial support. We acknowledge the DJEI/DES/SFI/HEA Irish Trinity Center for High-Performance Computing (TCHPC) for the provision of computational resources.

## AUTHOR DECLARATIONS

### Conflict of Interest

The authors have no conflicts to disclose.

### Author Contributions

**S. Hazra**: Formal analysis (equal); Investigation (equal); Methodology (equal); Validation (equal); Writing – original draft (equal); Writing – review & editing (equal). **U. Patil**: Formal analysis (equal);

Investigation (equal); Methodology (equal); Project administration (equal); Software (equal); Supervision (equal); Validation (equal); Writing – review & editing (equal). **S. Sanvito**: Conceptualization (lead); Formal analysis (equal); Funding acquisition (lead); Investigation (equal); Methodology (equal); Project administration (equal); Resources (lead); Supervision (equal); Writing – review & editing (lead).

## DATA AVAILABILITY

The data used to train and test two different models, at 100 and 300 K, are openly available in cZenodo at https://doi.org/10.5281/zenodo.15685946. The Python plugin module used to compute the Hartree potential will be available upon request. The JLCDM is openly available on Zenodo at https://doi.org/10.5281/zenodo.17197727.

## APPENDIX A: CARTESIAN PRODUCTS

In the method section, we have shown that the product of two Cartesian Gaussians, centered at different positions, **A** and **B**, can be written as the product of the three independent Cartesian components. These products comprise a polynomial and a Gaussian, now centered at a new position **P**. Equations (6) and (7) write such a polynomial in terms of the vector **r** − **P**. Here, we show that derivation explicitly,

$$
\begin{aligned}
x_A^{l_1} x_B^{l_2} &= (x-A_x)^{l_1}(x-B_x)^{l_2} \\
&= [(x-P_x)+(P_x-A_x)]^{l_1}[(x-P_x)+(P_x-B_x)]^{l_2} \\
&= (x_p-\overline{\mathbf{PA}}_x)^{l_1}(x_p-\overline{\mathbf{PB}}_x)^{l_2} \\
&= \sum_{i=0}^{l_1} x_P^i(\overline{\mathbf{PA}})_x^{(l_1-i)} C(l_1,i) \sum_{j=0}^{l_2} x_P^j(\overline{\mathbf{PB}})_x^{(l_2-j)} C(l_2,j) \\
&= \sum_{k=0}^{l_1+l_2} x_P^k \left[\sum_{i=0}^{k}\sum_{i=0}^{l_2} x_P^i x_P^j (\overline{\mathbf{PA}})_x^{(l_1-i)} C(l_1,i)(\overline{\mathbf{PB}})_x^{(l_2-j)} C(l_2,j)\right] \\
&= \sum_{k=0}^{l_1+l_2} x_P^k f_k(l_1,l_2,\overline{\mathbf{PA}}_x,\overline{\mathbf{PB}}_x),
\end{aligned}
\tag{A1}
$$

where $x_p = (x-P_x)$, $C(l_1,i) = \frac{l_1!}{i!(l_1-i)!}$, and $\overline{\mathbf{PA}}_x = (P_x - A_x)$. The binomial expansion, in fact, writes $(x_p - \overline{\mathbf{PA}}_x)^{l_1} = \sum_{i=0}^{l_1} x_P^i (\overline{\mathbf{PA}})_x^{(l_1-i)} C(l_1,i)$.

## APPENDIX B: EVALUATION OF THE $I'(\mathbf{r})$ INTEGRAL

Here, we provide a more explicit derivation of Eq. (16),

$$
\begin{aligned}
I'(\mathbf{r}') &= \int_0^\infty \frac{s^{L-2L'} s^{-1/2} e^{-\frac{\gamma s|\mathbf{r}'-\mathbf{P}|^2}{(\gamma+s)}}}{(\gamma+s)^{L-L'}(\gamma+s)^{3/2}} ds \\
&= \int_0^\infty \left(\frac{s}{\gamma+s}\right)^{L-L'} \frac{s^{-L'} s^{-1/2}}{(\gamma+s)^{3/2}} e^{-\frac{\gamma s|\mathbf{r}'-\mathbf{P}|^2}{(\gamma+s)^{3/2}}} ds.
\end{aligned}
\tag{B1}
$$

Now, by substituting $t^2 = \frac{s}{(\gamma+s)}$, $ds = \frac{2}{\gamma}(\gamma+s)^{3/2}s^{1/2}dt$, and $s = \frac{\gamma t^2}{(1-t^2)}$, the integral becomes

$$
I'(\mathbf{r}') = \frac{2}{\gamma^{(L'+1)}} \int_0^1 t^{2(L-2L')}(1-t^2)^{L'} e^{\gamma t^2|\mathbf{r}'-\mathbf{P}|^2} dt. \tag{B2}
$$

Again, by using the binomial expansion, the term $(1-t^2)^{L'}$ can be expanded as

$$
(1-t^2)^{L'} = \sum_{h=0}^{L'} (-1)^h \binom{L'}{h} t^{2h}. \tag{B3}
$$

Hence, we obtain

$$
\begin{aligned}
I'(\mathbf{r}') &= \sum_{h=0}^{L'} \frac{2(-1)^h}{\gamma^{L'+1}} \binom{L'}{h} \int_0^1 t^{2(L-2L'+h)} e^{-\gamma t^2|\mathbf{r}'-\mathbf{P}|^2} \\
&= \sum_{h=0}^{L'} \frac{2(-1)^h}{\gamma^{L'+1}} \binom{L'}{h} \int_0^{\gamma|\mathbf{r}'-\mathbf{P}|^2} u^{\left[(L-2L'+h)+\frac{1}{2}\right]-1} e^{-u} du \\
&= \sum_{h=0}^{L'} \frac{(-1)^h}{\gamma^{L-L'+h+1.5}} \times \binom{L'}{h} \\
&\quad \times \frac{\Gamma\left[L-2L'+h+0.5\right]\Gamma_{\text{inc}}\left[(L-2L'+h+0.5),\gamma|\mathbf{r}'-\mathbf{P}|^2\right]}{|\mathbf{r}'-\mathbf{P}|^{2(L-2L'+h)+1}}.
\end{aligned}
\tag{B4}
$$

Here, we have performed the substitution $\gamma t^2|\mathbf{r}'-\mathbf{P}|^2 = u$. The integral is finally expressed in term of the standard incomplete gamma function,

$$
\Gamma_{\text{inc}}(a,x) = \frac{1}{\Gamma(a)} \int_0^x t^{a-1} e^{-t} dt. \tag{B5}
$$